\documentclass{desyproc}

\begin{document}
\title{Search for low Energy solar Axions with CAST}

\author{{\slshape Giovanni Cantatore$^1$ for the CAST Collaboration$^2$}\\[1ex]
$^1$Universit\`a and INFN Trieste, Via Valerio 2, 34127, Trieste, Italy\\
\\
$^2${\it CAST Collaboration}\\
\footnotesize E.~Arik$^{17}$\footnote{Present address: Bogazici
University, Istanbul, Turkey}, S.~Aune$^{2}$,
D.~Autiero$^{1}$\footnote{Present address: Institute de Physique
Nucl\'eaire, Lyon, France}, K.~Barth$^{1}$, A.~Belov$^{10}$,
B.~Beltr\'an$^{5}$\footnote{Present address: Department of Physics,
Queen's University, Kingston, Ontario}, S.~Borghi$^{1}$,
F.~S.~Boydag$^{17}$, H.~Br\"auninger$^{4}$, G.~
Cantatore$^{18}$,J.~M.~Carmona$^{5}$, S.~Cebri\'an$^{5}$,
S.~A.~Cetin$^{17}$, J.~I.~Collar$^{6}$, T.~Dafni$^{2}$,
M.~Davenport$^{1}$, L.~Di~Lella$^{1}$\footnote{Present address:
Scuola Normale Superiore, Pisa, Italy},
O.~B.~Dogan$^{17}$\footnotemark[1], C.~Eleftheriadis$^{7}$, N.~Elias$^{1}$,
G.~Fanourakis$^{8}$, E.~Ferrer-Ribas$^{2}$, H.~Fischer$^{9}$,
J.~Franz$^{9}$, J.~Gal\'an$^{5}$, E. Gazis$^{20}$,
T.~Geralis$^{8}$, I.~Giomataris$^{2}$, S.~Gninenko$^{10}$,
H.~G\'omez$^{5}$, M.~Hasinoff$^{11}$, F.~H.~Heinsius$^{9}$,
I.~Hikmet$^{17}$, D.~H.~H.~Hoffmann$^{3}$,
I.~G.~Irastorza$^{2,5}$, J.~Jacoby$^{12}$,
K.~Jakov\v{c}i\'{c}$^{14}$, D.~Kang$^{9}$, T.
Karageorgopoulou$^{20}$, M.~ Karuza$^{18}$, K.~K\"onigsmann$^{9}$,
R.~Kotthaus$^{13}$, M.~Kr\v{c}mar$^{14}$, K.~Kousouris$^{8}$,
M.~Kuster$^{3}$, B.~Laki\'{c}$^{14}$, C.~Lasseur$^{1}$,
A.~Liolios$^{7}$, A.~Ljubi\v{c}i\'{c}$^{14}$, V.~Lozza$^{18}$,
G.~Lutz$^{13}$, G.~Luz\'on$^{5}$, D.~Miller$^{6}$,
A.~Morales$^{5}$\footnote{deceased}, J.~Morales$^{5}$,
T.~Niinikoski$^{1}$, A.~Nordt$^{3}$, A.~Ortiz$^{5}$,
T.~Papaevangelou$^{1}$, M.~Pivovaroff$^{16}$, A.~Placci$^{1}$,
G.~Raiteri$^{18}$, G.~Raffelt$^{13}$, H.~Riege$^{1}$,
A.~Rodr\'iguez$^{5}$, J.~Ruz$^{5}$, I.~Savvidis$^{7}$,
Y.~Semertzidis$^{15}$, P.~Serpico$^{13}$, S.~K.~Solanki$^{19}$,
R.~Soufli$^{16}$, L.~Stewart$^{1}$,M.~Tsagri$^{15}$,
K.~van~Bibber$^{16}$, J.~Villar$^{5}$, J.~Vogel$^{9}$,
L.~Walckiers$^{1}$, K.~Zioutas$^{1,15}$\\
\\
1. European Organization for Nuclear Research (CERN), Gen\`eve, Switzerland\\
2. DAPNIA, Centre d'\'Etudes Nucl\'eaires de Saclay (CEA-Saclay), Gif-sur-Yvette, France\\
3. Technische Universit\"{a}t Darmstadt, IKP, Darmstadt, Germany\\
4. Max-Planck-Institut f\"{u}r extraterrestrische Physik, Garching, Germany\\
5. Instituto de F\'{\i}sica Nuclear y Altas Energ\'{\i}as, Universidad de Zaragoza, Zaragoza, Spain\\
6. Enrico Fermi Institute and KICP, University of Chicago, Chicago, IL, USA\\
7. Aristotle University of Thessaloniki, Thessaloniki, Greece\\
8. National Center for Scientific Research ``Demokritos'', Athens, Greece\\
9. Albert-Ludwigs-Universit\"{a}t Freiburg, Freiburg, Germany\\
10. Institute for Nuclear Research (INR), Russian Academy of Sciences, Moscow, Russia\\
11. Department of Physics and Astronomy, University of British Columbia, Department of  Physics, Vancouver, Canada\\
12. Johann Wolfgang Goethe-Universit\"at, Institut f\"ur Angewandte Physik, Frankfurt am Main, Germany\\
13. Max-Planck-Institut f\"{u}r Physik (Werner-Heisenberg-Institut), Munich, Germany\\
14. Rudjer Bo\v{s}kovi\'{c} Institute, Zagreb, Croatia\\
15. Physics Department, University of Patras, Patras, Greece    \\
16. Lawrence Livermore National Laboratory, Livermore, CA, USA\\
17. Dogus University, Istanbul, Turkey\\
18. Instituto Nazionale di Fisica Nucleare (INFN), Sezione di
Trieste and Universit\`a di Trieste, Trieste, Italy\\
19. Max-Planck-Institut f\"{u}r Aeronomie, Katlenburg-Lindau, Germany\\
20. National Technical University of Athens, Athens, Greece\\
}

\contribID{cantatore\_giovanni\_lea}

\desyproc{DESY-PROC-2008-02}
\acronym{Patras 2008} 
\doi 

\maketitle

\begin{abstract}
We have started the development of a detector system, sensitive to single photons in the eV energy range, to be suitably coupled to one of the CAST magnet ports. This system should open to CAST a window on possible detection of low energy Axion Like Particles emitted by the sun.
Preliminary tests have involved a cooled photomultiplier tube coupled to the CAST magnet via a Galileian telescope and a switched 40 m long optical fiber. This system has reached the limit background level of the detector alone in ideal conditions, and two solar tracking runs have been performed with it at CAST. Such a measurement has never been done before with an axion helioscope. We will present results from these runs and briefly discuss future detector developments.\end{abstract}

\section{Introduction}
It has recently been pointed out \cite{zioutas_sun} that many phenomena taking place in the sun, especially in its corona and in its magnetic field, are far from being completely understood. The production by the sun of Axion Like Particles (ALPs) and their subsequent interactions in the solar environment could provide a key to interpreting the physical mechanisms underlying these phenomena.  These, and other considerations led to starting a search with the CAST magnetic helioscope \cite{cast} for hypothetical ALPs emitted by the sun in the energy range below 100 eV. The first step of this search, which will be reported here, has involved looking for 2-4 eV photons produced in the CAST magnet bore by the Primakoff \cite{primakoff} conversion into photons of solar ALPs in the latter energy range. The short term objective was to efficiently couple a detector system sensitive in the eV energy range to a CAST magnet bore and evaluate its background in normal operating conditions. The long term objective of the effort is attempting to detect, using sensors with the appropriate spectral sensitivity and good enough background, "low"-energy (tens of eV's) photons generated in the CAST helioscope by possible interactions of low-energy solar ALPs. 
We will briefly describe the detector system, which has been developed for this purpose under the BaRBE project financed by the Italian Istituto Nazionale di Fisica Nucleare (INFN), along with the coupling of this system to the CAST magnet. Finally, the data taking campaigns will be discussed and a summary of the data presented.

\section{Preliminary tests and system set-up}
The starting idea of the BaRBE project is to begin with readily available photon detectors sensitive in the visible range, test them in ideal laboratory conditions, and then design and build an optical system to couple the detectors to one of the bores of the CAST magnet. The devices used in this initial phase were a photomultiplier tube (PMT) (model 9893/350B made by EMI-Thorn) and an avalanche photodiode (APD) (model id100-20 made by idQuantique). The PMT had an active area dia. of 9 mm, with peak sensitivity at 350 nm. The APD had an active area dia. of  20 $\mu$m and peak sensitivity at 500 nm. For both detectors the Dark Count Rate (DCR) measured in the laboratory during preliminary tests was about 0.4 Hz.
To measure the DCR, the PMT was biased at 1950 V and instrumented with an electronic readout chain consisting of NIM standard modules. The APD readout module gave a 2 V, 10 ns wide, output signal which was trasformed into a TTL pulse via a custom circuit. Both detectors were cooled by means of their built in Peltier coolers, operating at -20 $^o$C, and were illuminated by a suitably attenuated blue LED source. To demonstrate single photon operation, the count histogram for each detector was fitted with a Poissonian distribution with average equal to 1. The DCR  was then obtained after cutting the electronic noise pedestal using the fitted curve.


\begin{figure}[hb]
\centerline{\includegraphics[width=0.75\textwidth]{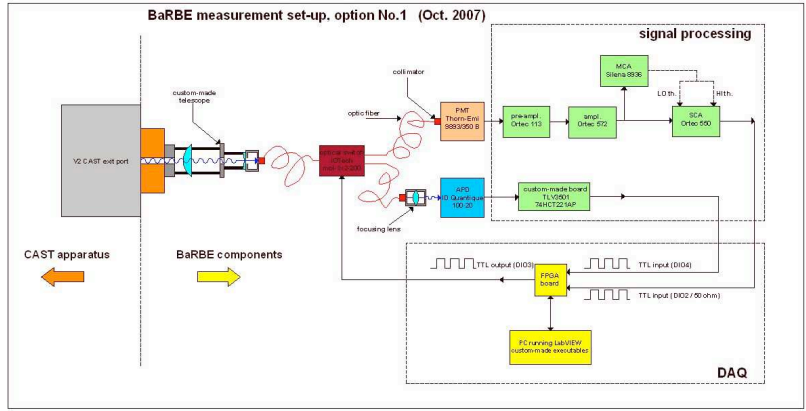}}
\caption{Schematic block-diagram of the BARBE setup at CAST (see text).}\label{Fig:figure1}
\end{figure}

The basic elements of the coupling system were a Galileian telescope, a 40 m multimode optical fiber complete with input collimator and an optical switch (mod. 1x2 made by Leoni). The Galileian telescope consisted of a 2 inch dia., f = 200 mm, convex lens and of a 1 inch dia., f = - 30 mm, concave lens and was designed to optically couple the 40 mm dia. bore of the CAST magnet into the 9 mm dia. input collimator of the multimode fiber. The telescope was mounted directly onto one of the sunset side ports of the CAST magnet \footnote{The telescope optical axis, determined before installation on a separate bench using an auxiliary laser beam, was aligned to the CAST magnet axis as established by the surveyors.}. The detectors were placed far away from the magnet in the CAST experimental hall. The optical switch, which can be triggered by external TTL pulses, was used in order to share the light coming from the fiber between the PMT and the APD: each detector could then look at the magnet bore for 50\% of the time and at the background for the remaining 50\%. Figure \ref{Fig:figure1} shows a block-diagram of the layout of the system as mounted on CAST.
The overall light collection efficiency was about 50\% for the PMT and less than 1\% for the APD. The low efficiency of the APD channel was due to unresolved focussing difficulties. Since the APD data are of inferior quality, only the PMT measurements are considered. The total number of counts in each measurement is affected by afterpulses generated either by the PMT itself or by its readout electronic chain. It was found that afterpulses account for 11\% of total counts.
To eliminate the effect of the afterpulses the mean rate of counts is calculated, fot both light and dark counts by solving for $x = 0$ the equation $N_{x} = A \cdot e^{-m}m^{x}/x!$, where x is the channel number, A is the total number of occurences in all channels, $N_{x}$ is the number of occurences in the x-th channel measured experimentally. In this way occurrences in channel 0 are not affected by afterpulses.

\section{Measurement results}
Two measurement campaigns were conducted. The first one in November 2007 with the telescope attached to the V2 port of the CAST magnet, while the second one was in March 2008 with the telescope on the V1 port. 

\begin{wrapfigure}{r}{0.75\textwidth}
\centerline{\includegraphics[width=0.65\textwidth]{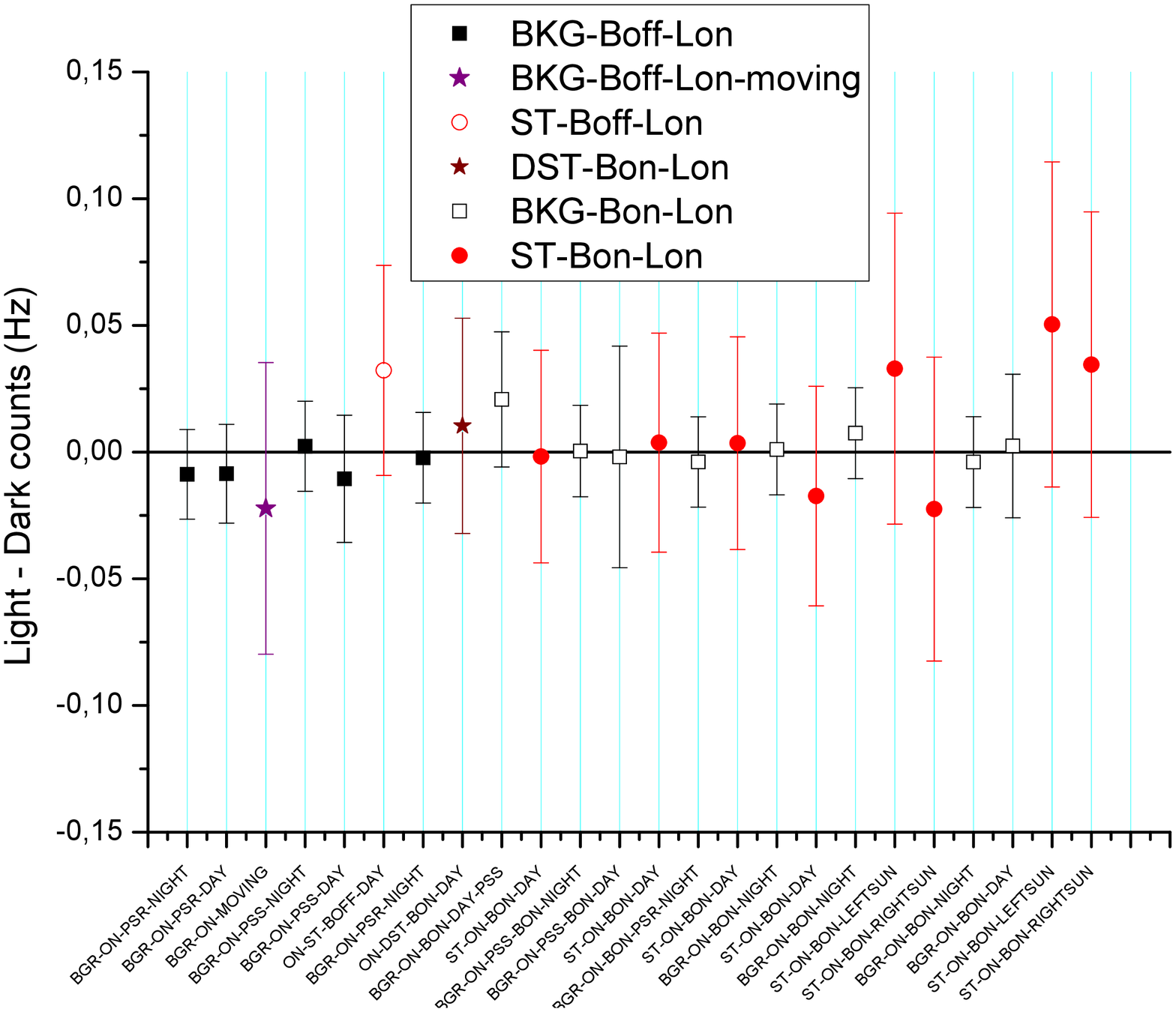}}
\caption{Difference between "light" and "dark" average count rates measured in the March 2008 run (PMT data only). Points on the abscissa axis correspond to different configurations of the entire apparatus, including background (BKG) tests and sun tracking (ST) periods with the magnet on (Bon).  Error bars represent 1 $\sigma$ intervals (see also text).}\label{Fig:figure2}
\end{wrapfigure}

In the first run each detector, PMT and APD, looked at the magnet bore for 50\% of the time and at the background for the other 50\%. Environmental checks and background measurements in different magnet positions and with field on and off were performed. A total of 45000 s of "live" data with the magnet on were taken, 10000 s of dummy solar tracking data and 35000 s of actual sun tracking data.

The second run was conducted using the PMT only, but keeping the switching system in order to have again the detector share its live time equally between signal and background. In this case 45000 s of live data were taken, of which 5000 s of dummy solar tracking data, 20000 s of actual solar tracking data and 20000 s of data with the magnet pointing off the sun center (10000 s pointing 0.25$^o$ to the right and 10000 s pointing 0.25$^o$ to the left). Figure \ref{Fig:figure2} shows a plot of the difference of the measured average count rates between "light", when the PMT was looking at the magnet bore, and  "dark", when the optical switch was toggled on the other position. The abscissa axis refers to the different conditions in which data were taken. Taking into account the 1 $\sigma$ error bars reported in the plot, no statistically significant difference is found between "light" and "dark" count rates. The average background count rate measured for 3-4 eV photons during solar tracking was 0.35 $\pm$ 0.02 Hz for a total of 75000 s of data.

\section{Conclusions and perspectives}
The measurement runs conducted with the BaRBE detector system demonstrated that it is possible to couple two detectors to the CAST magnet via an optical fiber, while preserving a reasonable light collection efficiency (50\% in the PMT case, that corresponds to 10\% of overall system efficiency when taking into account the PMT spectral response curve), and without introducing additional noise sources.
In 12 data sets taken during solar tracking (including 2 sets pointing off center) the background count rate for 3-4 eV photons was 0.35 $\pm$ 0.02 Hz and no significative excess counts over background were observed. This is the first time such a measurement has been done with an axion helioscope.

The challenge is now to progress to a new detector(s) with lower intrinsic background, possibly extending the spectral sensitivity to other regions of the energy interval below 100 eV . In the case of visible photons, one could also envision enclosing the CAST magnet bore in a resonant optical cavity in order to enhance the axion-photon conversion probability \cite{axion_photon}. This would however require solving rather complex compatibility problems with the rest of the CAST apparatus.

Three
types of detectors have at this moment been considered for future developments, a Transition Edge Sensor (TES) \cite{tes}, a silicon sensor with DEPFET readout \cite{depfet} and an APD cooled to liquid nitrogen temperatures. The TES sensor promises practically zero background, spectroscopic capability and sensitivity from less than 1 eV up to tens of eV's. It however requires operation at 100 mK and it has a small sensitive area (about $100 \mu \mbox{m} \times 100 \mu \mbox{m}$). The DEPFET sensors could reach a very low background if used in the Repetitive Non Destructive Readout (RNDR) mode, however they also have a small sensitive area. Finally, the cooled APD could be operated relatively easily if one accepts afterpulsing events, which should not pose a problem in a low expected rate environment. On the other hand, the sensitive area would only be about 300 $\mu \mbox{m}^2$ and the spectral sensitivity limited to the visible region.
The present plan is to initially pursue all three possibilities in the hope of identifying the one where progress is faster.

\section{Acknowledgments}
Funding for the BaRBE project has been provided by INFN (Italy). The support by the GSRT in Athens is gratefully acknowledged. This research was partially supported by the ILIAS project funded by the EU under contract EU-RII3-CT-2004-506222. Special thanks to M. Karuza, V. Lozza and G. Raiteri of INFN Trieste for their excellent detector work, and to the entire CAST Collaboration for wonderful support and assistance. We also warmly thank the CERN staff for their precious help.


\begin{footnotesize}



%

\end{footnotesize}


\end{document}